\documentclass[final]{raa06}           
\usepackage{graphicx,times}             
\usepackage{natbib}
\usepackage{amssymb,amsmath}
\bibpunct{(}{)}{;}{a}{}{,}
\usepackage[colorlinks=true, citecolor=blue]{hyperref}%

\usepackage{graphicx,kantlipsum,setspace}
\usepackage{caption}
\usepackage{graphics,epsf}
\usepackage{amsmath}                
\usepackage{amsfonts}               
\usepackage{amssymb}                
\usepackage{epsfig}                 
\usepackage{appendix}
\usepackage{graphicx}
\usepackage{float}
\usepackage{color}
\usepackage{multirow}
\usepackage{colortbl}
\usepackage[para,online,flushleft]{threeparttable}
\usepackage{xcolor}

\hypersetup{citecolor=blue, 
            linkcolor=red, 
            menucolor=blue, 
            urlcolor=blue}  

\newcommand{\s}{{~\rm s}}

\newcommand{\erg}{{~\rm erg}}
\newcommand{\yr}{{~\rm yr}}

\begin{document}

   \title{On the response of massive main sequence stars to mass accretion and outflow at high rates
}

   \volnopage{Vol.0 (20xx) No.0, 000--000}      
   \setcounter{page}{1}          

   \author{Ealeal Bear, Noam Soker
    }

   \institute{Department of Physics, Technion, Haifa, 3200003, Israel;   {\it   ealeal44@technion.ac.il; soker@physics.technion.ac.il}\\
\vs\no
   {\small Received~~20xx month day; accepted~~20xx~~month day}}

\abstract{With a one-dimensional stellar evolution model, we find that massive main-sequence stars can accrete mass at very high mass accretion rates without expanding much if they lose a significant fraction of this mass from their outer layers simultaneously with mass accretion. We assume the accretion process is via an accretion disk that launches powerful jets from its inner zones. These jets remove the outer high-entropy layers of the mass-accreting star. This process operates in a negative feedback cycle, as the jets remove more envelope mass when the star expands. With the one-dimensional model, we mimic the mass removal by jets by alternative mass addition and mass removal phases. For the simulated models of $30 M_\odot$ and $60 M_\odot$, the star does not expand much if we remove more than about half of the added mass in not-too-short episodes. This holds even if we deposit the energy the jets do not carry into the envelope. As the star does not expand much, its gravitational potential well stays deep, and the jets are energetic.  These results are relevant to bright transient events of binary systems powered by accretion and the launching of jets, e.g., intermediate luminosity optical transients, including some luminous red novae, the grazing envelope evolution, and the 1837-1856 Great Eruption of Eta Carinae.
 }

 \authorrunning{E. Bear, N. Soker}            
\titlerunning{On the response of massive main sequence stars}  
   
      \maketitle

\section{Introduction}
\label{sec:intro}

Intermediate luminosity optical transients (ILOTs; \citealt{Berger2009, KashiSoker2016Terms, MuthukrishnaetalM2019}) are a heterogeneous group of transients in the visible band with typical peak luminosities between those of classical novae and those of supernovae (e.g., \citealt{Mouldetal1990, Bondetal2003, Rau2007, Ofek2008, Masonetal2010, Kasliwal2011, Tylendaetal2013, Kasliwaletal2012, Kaminskietal2018, BoianGroh2019, Caietal2019, Jencsonetal2019, Kashietal2019Galax, PastorelloMasonetal2019, Blagorodnovaetal2020, Banerjeeetal2020, Howittetal2020, Jones2020, Kaminskietal2020Nova1670, Kaminskietal2021Nova1670, Klenckietal2021, Stritzingeretal2020AT2014ej, Stritzingeretal2020SNhunt120, Blagorodnovaetal2021, Mobeenetal2021, Pastorelloetal2021, Pastorelloetal2023, Addisonetal2022, Caietal2022,  Wadhwaetal2022, Kaminskietal2023, Karambelkaretal2023,  ZainMobeenetal2024, Kaminski2024}).
There is no consensus on the naming of this heterogeneous transient group. \textit{We use the term ILOTs for all transients powered by gravitational energy,} whether triggered by a merger process or a mass transfer. Other researchers (e.g., \citealt{Jencsonetal2019, Caietal2022b})  use other terms, like gap transients, luminous red novae (LRNe), red novae, or intermediate luminosity red transients. There is also no consensus on the sub-classes of this group of transients (e.g., \citealt{KashiSoker2016Terms} versus \citealt{PastorelloMasonetal2019} and \citealt{PastorelloFraser2019}). 
We use the term LRNe for ILOTs that are powered by a complete merger that leaves one stellar remnant \citep{KashiSoker2016Terms}, and for a common envelope evolution (CEE) as during the bright phase there is only one photosphere. We also include ILOTs from the grazing envelope evolution of low-mass stars (to distinguish from eruptions of luminous blue variables) under LRNe.  

CEE LRNe might be powered even by a sub-stellar companion, i.e., a planet or a brown dwarf (e.g., \citealt{RetterMarom2003,  Metzgeretal2012, Yamazakietal2017, Kashietal2019Galax, Gurevichetal2022, Deetal2023, Oconnoretal2023}), although most events are thought to be powered by a stellar companion (e.g., \citealt{Tylendaetal2011, Ivanovaetal2013a, Nandezetal2014, Kaminskietal2015, Pejchaetal2016a, Pejchaetal2016b, Soker2016GEE, Blagorodnovaetal2017, MacLeodetal2017, MacLeodetal2018,  Segevetal2019, Howittetal2020, MacLeodLoeb2020, Qianetal2020, Schrderetal2020, Blagorodnovaetal2021, Addisonetal2022, Zhuetal2023, Tylendaetal2024}). In CEE LRNe, the primary energy sources might be the dynamical interaction of the companion inside the envelope of the engulfing larger star, the recombination of the ejected common envelope (e.g., \citealt{MatsumotoMetzger2022}), or the accretion energy onto the more compact companion. 
The dynamical interaction is the gravitational energy of the companion that spirals in and ejects the envelope. Radiation can result from the collision of ejected envelope gas with itself in and near the equatorial plane (e.g., \citealt{Pejchaetal2016a, Pejchaetal2016b, Pejchaetal2017, MetzgerPejcha2017, HubovaPejcha2019}), or from the heating of the envelope by the spiraling-in companion. 

We adopt the view that the most efficient energy source to power a bright ILOT is the accretion energy of gas onto one of the two stars of a binary system followed by the launching of jets (e.g., \citealt{Soker2020ILOTjets}). The bipolar ejecta of spatially-resolved ILOTs (e.g., \citealt{Kaminski2024}) suggest that most, or even all, bright ILOTs are powered by jets (e.g.,  \citealt{Soker2023BrightILOT, Soker2024}). The companion accretes mass via an accretion disk and launches jets. When the companion is a neutron star or a black hole, the event can mimic a core-collapse supernova (e.g., \citealt{SokerGilkis2018, Gilkisetal2019, GrichenerSoker2019, YalinewichMatzner2019, Schreieretal2021}), and the event is termed common envelope jet supernova rather than an LRN. In a CEE LRN (including an LRN in a grazing envelope evolution; \citealt{Soker2016GEE}), the companion launches jets that collide with the common envelope or with the circumstellar material (CSM), a process that transfers kinetic energy to thermal energy and radiation (e.g., \citealt{Soker2020ILOTjets, SokerKaplan2021RAA}). 

Luminous blue variables are also a group of ILOTs. The most famous is the Great Eruption of Eta Carinae, which ejected a bipolar nebula, the Homunculus. The bipolar morphology of the Homunculus testifies that jets powered the Great Eruption of Eta Carinae (e.g., \citealt{Soker2001, KashiSoker2010}). The mass of the companion that accreted the gas during the great eruption and launched the jets is $M_2 \simeq 30- 80 M_\odot$ \citep{KashiSoker2016EtaCar}. 
{ \cite{KashiSoker2010} estimated the average mass accretion rate onto the companion during the 20-year-long Great Eruption as $\simeq 0.2 M_\odot \yr^{-1}$. We aim here to explain such high mass accretion rates without envelope expansion. Due to numerical difficulties, we will not simulate such high rates, but we will present the principles of the process. }
We will examine the response of such a companion to mass accretion accompanied by mass removal that we attribute to jet launching.  

{ The large momentum of some planetary nebulae and pre-planetary nebulae also suggest that the companion can accrete mass at a high rate and launch powerful jets (e.g., \citealt{BlackmanLucchini2014}). Most of the companions to the central stars of planetary nebulae are low-mass main sequence stars, for which \cite{BlackmanLucchini2014} found the required accretion rate in these high-momentum planetary nebulae to be $\approx 10^{-3} M_\odot \yr^{-1}$. The stars we simulate in this study are two orders of magnitude more massive, so the accretion rate is scaled to $\approx 0.1 M_\odot \yr$. }
 
Another group of ILOTs includes pre-explosion outbursts, where a massive star experiences an outburst years to days before a core-collapse supernova explosion. The energy source of these outbursts can be accretion onto a companion (e.g., \citealt{McleySoker2014, DanieliSoker2019, Tsunaetal2024}; see \citealt{Soker2022Rev} for a review). The compact companion accretes mass and launches jets that power the pre-explosion outburst when colliding with the CSM. We here consider cases where the compact object is a main sequence companion with a mass of $M_2 \ga 1.5 M_\odot$ and therefore has a radiative envelope.   

In this study, we examine one aspect of ILOTs powered by the accretion process onto massive main sequence stars, i.e., those with a radiative envelope. We further consider that the mass-accreting main sequence star launches jets. We will not study the entire accretion via an accretion disk that launches jets but rather mimic the process with the numerical spherical code \textsc{mesa} (Section \ref{sec:Method}). We present our results in Section \ref{sec:Results}. In Section \ref{sec:Summary} we summarize our results, compare them to a study by \cite{SchurmannLanger2024}, and discuss their implications to ILOTs and CEE with massive main sequence companions. 

\section{Method}
\label{sec:Method}
We used version 23.05.1 of the stellar evolution code Modules for Experiments in Stellar Astrophysics (\textsc{mesa}; \citealt{Paxtonetal2011, Paxtonetal2013, Paxtonetal2015, Paxtonetal2018, Paxtonetal2019, Jermynetal2023}) in its single star mode. We evolve a stellar model with an initial zero-age main sequence (ZAMS) mass of $M=30M_\odot$ and $M=60M_\odot$ and metallicity of $z=0.019$. We start mass accretion and energy deposition on the main sequence at the age of $t_{\rm MS}= 1.3 \times 10^6 \yr$ in all simulations of $M=30M_\odot$ and $t_{\rm MS}= 0.9 \times 10^6 \yr$ in all simulations of $M=60M_\odot$. 
We base our simulations on the example of $\textit{20M~pre~ms~to~core~collapse}$ for both $M=30M_\odot$ and $M=60M_\odot$; all other parameters remain as in the default of \textsc{mesa}.\footnote{  
The default capabilities of \textsc{mesa}-single relay on the MESA EOS that is a blend of the OPAL \citep{RogersNayfonov2002}, SCVH
\citep{Saumonetal1995}, FreeEOS \citep{Irwin2004}, HELM \citep{TimmesSwesty2000}, PC \citep{PotekhinChabrier2010}, and Skye \citep{Jermynetal2021} EOSes. Radiative opacities are primarily from OPAL \citep{IglesiasRogers1993, IglesiasRogers1996}, with low-temperature data from \citet{Fergusonetal2005} and the high-temperature, Compton-scattering dominated regime by
\citet{Poutanen2017}.  Electron conduction opacities are from
\citet{Cassisietal2007} and \citet{Blouinetal2020}.
Nuclear reaction rates are from JINA REACLIB \citep{Cyburtetal2010}, NACRE \citep{Anguloetal1999} and additional tabulated weak reaction rates \citealt{Fulleretal1985, Odaetal1994, Langankeetal2000}.  Screening is included via the prescription of \citet{Chugunovetal2007}. Thermal neutrino loss rates are from \citealt{Itohetal1996}.}

We mimic the process by which the jets remove envelope mass from the mass-accreting star using small alternating pulses of accretion and mass removal; each mass-addition part has a duration of $\Delta t_{\rm p}$. The code \textsc{mesa} removes and adds mass with the same properties, like entropy, as in the outermost shell of the stellar model. 
Our pulse has two parts. In the first part, when we accrete mass, we also deposit energy. In the second part of the pulse, we remove a mass, either $\eta_{\rm MR}=\frac{2}{3}$ or $\eta_{\rm MR}=\frac{1}{3}$ of the mass accreted in the first part of the pulse. We deposit energy to the envelope's outer $0.1$ (by radius). 
We found (see Section \ref{sec:Results}), that if we remove a small fraction, which for many simulations is $\eta_{\rm MR}=\frac{1}{3}$ of the accreted mass, the star inflates to a very large radii, that violet the assumption of the model, like hydrostatic equilibrium. 
In all simulations, we start the pulses on the early main sequence. The exact time at the main sequence, when we start the simulations, has little influence on the results { because only the outer envelope participates in the mass addition and removal, which changes little during the main sequence.  }

The amount of added energy is $\eta_{\rm acc}=0.25$ of the accreted energy during the mass addition phase. { In most simulations that fulfill our goal of non-expansion (moderate in real situations by feedback), the mass we remove in a cycle is $2/3$ of the mass we add. The energy required to remove this mass equals the energy this mass deposited while accreted. However, the removed mass will reach a positive terminal velocity, implying that it carries more energy than it added while accreted. Therefore, the energy left in the stellar envelope is $\eta_{\rm acc}<0.33$ of the accreted energy. Since the ejected mass will have a terminal speed close to the escape speed, we expect this fraction to be $\eta_{\rm acc} \ll 0.33$. We take a conservative approach of $\eta_{\rm acc}=0.25$; taking an even lower value, as we expect, will result in more moderate envelope expansion, which will act in favor of our scenario.     } 

{ Following mass addition and removal and energy injection, the star either rapidly expands or reaches a more or less constant radius (or only expands very slowly). We stop the simulations when we identify the response of the star.  }

\section{Results}
\label{sec:Results}

We aim to present a process by which massive main sequence stars can accrete mass at high rates without expanding much. We focus, therefore, on presenting the evolution of the stellar radius following a high mass accretion rate under different conditions. We emphasize that the process we are mimicking is of accretion from an accretion disk and a mass loss by jets launched from the inner zone of the accretion disk and its boundary with the stellar surface. The jets remove more material from the stellar outskirts. The inflow-outflow occurs simultaneously. However, we alternate mass accretion and mass loss because of the limitations of the one-dimensional stellar model. We refer to each mass addition phase as a pulse of duration $\Delta t_{\rm p}$. During the mass addition pulse, we also inject energy into the envelope. After this pulse, we remove a mass. In most cases, the duration of mass removal is equal to that of mass addition (the pulse duration). In two simulations, the mass removal time is longer. 

To avoid complicated graphs, the first three figures present only the mass addition and energy deposition parts of the pulses but not the mass removal parts. This causes a discontinuity in the lines from one pulse to the next. Namely, after the mass removal phase, the star contracts, and its mass decreases; therefore, the following line segment (next pulse) starts below and to the left of the endpoint of the line segment of the previous pulse.  

\begin{figure}[t]
	\centering
\includegraphics[trim=3.2cm 8.4cm 3.5cm 8.5cm ,clip, scale=0.60]{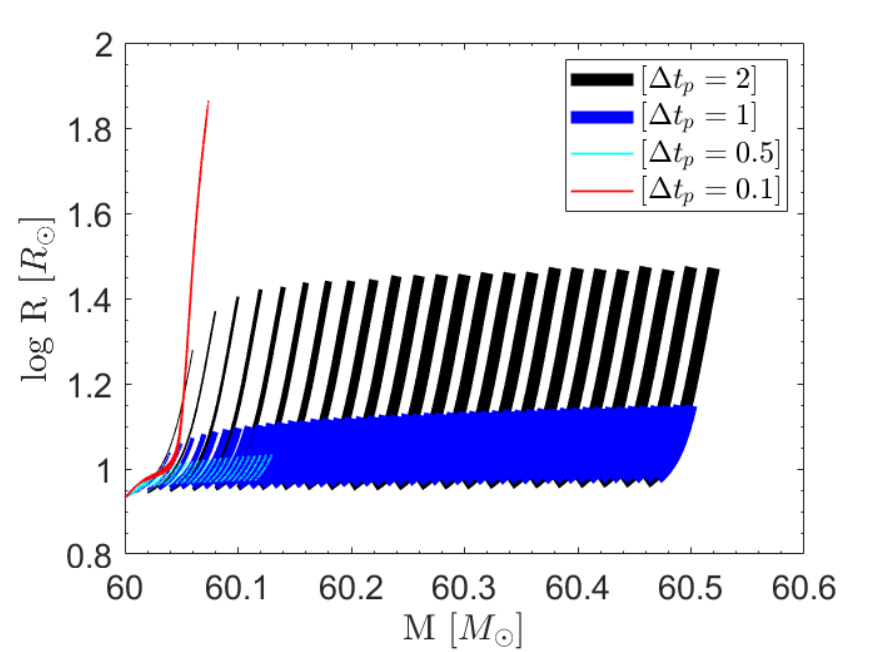}
\caption{Stellar radius (on a log scale) vs. stellar mass of a main sequence star with $M_{\rm ZAMS}=60M_\odot$, and for four different mass addition pulse durations, as follows. We present only the evolution during mass addition (pulses) but not during mass removal; this causes discontinuity in each line.  
Black line (total of 24 pulses) is for a pulse duration of  $\Delta t_{\rm p} = 2\yr$,  blue line (48 pulses) is for $\Delta t_{\rm p}=1\yr$, cyan line (24 pulses) represents pulses of $\Delta t_{\rm p}=0.5\yr$, and the red line (72 pulses) represents pulses of $\Delta t_{\rm p}=0.1\yr$. 
For all pulses, the mass accretion rate when we add mass is $\dot M_{\rm add} =0.03 M_\odot \yr^{-1}$. The mass removal rate when we remove mass is $\dot M_{\rm rem} = -0.02 M_\odot \yr^{-1}$, i.e., $\eta_{\rm MR} =\vert \dot M_{\rm rem} \vert /  \dot M_{\rm add} =0.67$, the mass addition and removal times are equal, and the net mass accretion rate is $\dot M_{\rm acc}= 0.005 M_\odot \yr^{-1}$. We add energy to the outer $10 \%$, of the radius of the stellar envelope when we add mass at a power of $\dot E_{\rm add}=6.32\times 10^{39} \erg \s^{-1}$, which is a fraction of $\eta_{\rm acc}=0.25$ of the gravitational energy that the accreted mass releases (see Section \ref{sec:Method}). The net energy deposition power (as we add energy only during half the cycle) is $E_{\rm acc}=3.16\times 10^{39} \erg \s^{-1}$. 
}
\label{fig:60Mo_logR_M}
\end{figure}

Figure \ref{fig:60Mo_logR_M} shows the evolution of the stellar radius of a main sequence star with $M_{\rm ZAMS} = 60 M_\odot$ for different pulse durations as indicated in the inset and caption. The total number of pulses is different between the simulations (see caption), as we stop the simulations after we identify the stellar behavior, namely if it rapidly expands, if it reaches a more or less constant radius, or if only a very slow expansion when we add more mass.  In the four cases of Figure \ref{fig:60Mo_logR_M}, we add mass and energy at rates of $\dot M_{\rm add} =0.03 M_\odot \yr^{-1}$ and $\dot E_{\rm add}=6.32\times 10^{39} \erg \s^{-1}$, respectively, and remove mass at a rate of $\dot M_{\rm rem} = -0.02 M_\odot \yr^{-1}$. The net average mass accretion rate and energy power are $\dot M_{\rm acc}= 0.005 M_\odot \yr^{-1}$ and $E_{\rm acc}=3.16\times 10^{39} \erg \s^{-1}$, respectively. The power of the accretion is a fraction of $\eta_{\rm acc}=0.25$ of the gravitational energy that the accreted mass releases (see Section \ref{sec:Method}). 

In three cases of Figure \ref{fig:60Mo_logR_M}, the star does not expand much; in one case, it rapidly expands. The four simulations of Figure \ref{fig:60Mo_logR_M} have $\eta_{\rm MR} =\vert \dot M_{\rm rem} \vert /  \dot M_{\rm add} =0.67$. We found that when $\eta_{\rm MR} = 0.33$, i.e., when we remove less mass after a mass addition pulse, the star rapidly expands in all cases, namely, for all pulse durations as in Figure \ref{fig:60Mo_logR_M}. This shows that to prevent rapid stellar expansion, the star must lose its outer high-entropy layers. According to our assumption, the jets from the accretion disk carry high-entropy gas, remove energy, and remove high-entropy gas from the envelope outskirts. We mimic this process. 

\begin{figure}[t]
	\centering
\includegraphics[trim=3.2cm 8.4cm 3.5cm 8.5cm ,clip, scale=0.54]{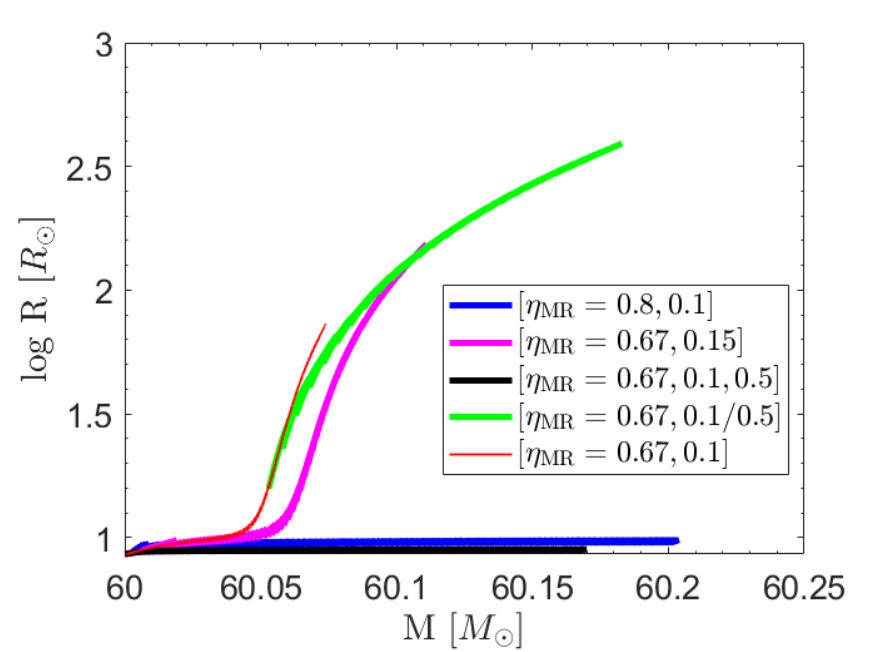}
\caption{Similar to Figure \ref{fig:60Mo_logR_M} but for more cases. 
The thin-red line (72 pulses) is the same as the thin-red line in Figure \ref{fig:60Mo_logR_M}, i.e., mass addition pulse lasts for $\Delta t_{\rm p}=0.1\yr$ and the mass removal rate is $\eta_{\rm MR}=0.67$ of the mass addition rate and last for the same length of $0.1 \yr$. The magenta line represents the same $\eta_{\rm MR}=0.67$ but for a pulse duration of $\Delta t_{\rm p}=0.15\yr$ (72 pulses).
The green line (24 pulses) is for a variation of the parameters where we let the star evolve on the red track. i.e., $\Delta t_{\rm p}=0.1\yr$ when $\eta_{\rm MR}=0.67$  (for 55 pulses) and when the stellar radius is $16 R_\odot$ we switch to $\Delta t_{\rm p}=0.5\yr$ at $\eta_{\rm MR}=0.67$ (for 24 additional pulses). 
The blue line represents simulation with $\Delta t_{\rm p}=0.1\yr$ but with a larger mass removal rate such that $\eta_{\rm MR}=0.8$. In this case, the star does not expand (336 pulses). The black line (168 pulses) represents a simulation with $\Delta t_{\rm p}=0.1\yr$ and mass addition rate of $\dot M_{\rm add}=0.03 M_\odot \yr^{-1}$, but the mass removal rate is $\dot M_{\rm rem}=-0.004 M_\odot \yr^{-1}$ and it lasts for $5 \Delta t_{\rm p}=0.5\yr$; this implies an effective value of $\eta_{\rm MR}=0.67$. In this case, the net average mass accretion rate is $\dot M_{\rm acc} = 0.00167 M_\odot \yr^{-1}$; the star does not expand. 
For all simulations in this figure, the power at which we add energy to the envelope during the mass addition phase (the pulse) is $\dot E_{\rm add} = 6.32\times 10^{39} \erg \s ^{-1}$ which is equal to $\eta_{\rm acc}=0.25$ accreted energy and is inserted in the outer $10\%$ of the star (by radius; see text).
}
\label{fig:60Mo_logR_M_0_1variation}
\end{figure}

Figure \ref{fig:60Mo_logR_M} shows that when the pulse duration is short, the star rapidly expands (red line for $\Delta t_{\rm p}=0.1 \yr$). We present more simulated cases in Figure \ref{fig:60Mo_logR_M_0_1variation} to investigate this phenomenon.
The thin-red line presents the same case as the thin-red line in Figure  \ref{fig:60Mo_logR_M}. 

Firstly, we note from Figure \ref{fig:60Mo_logR_M_0_1variation} that even for $\Delta t_{\rm p} = 0.15 \yr$, keeping other parameters as in the cases of Figure \ref{fig:60Mo_logR_M}, the star rapidly expands (the thick magenta line). However, if we remove more mass during the mass removal phase, the star does not expand much, as the blue line shows for the case where the mass removal rate is $\eta_{\rm MR} = 0.8$ of the mass accretion rate; namely, the new average mass accretion rate is $\dot M_{\rm acc}= 0.003 M_\odot \yr^{-1}$. This simulation was run for 336 pulses.
The green line in Figure \ref{fig:60Mo_logR_M_0_1variation} shows that once the star rapidly expands, it is difficult to halt this expansion. The green line represents a simulation that starts like the thin-red line, with $\Delta t_{\rm p} =0.1\yr$ and $\eta_{\rm MR}=0.67$. Then, when the star has expanded as a result of mass accretion to a radius of $16 R_\odot$ we switch to $\Delta t_{\rm p}=0.5\yr$, keeping $\eta_{\rm MR}=0.67$. The star continues to expand, although at a somewhat slower rate. We recall that when we start with $\Delta t_{\rm p}=0.5\yr$ the star does not expand (cyan line in figure \ref{fig:60Mo_logR_M}).  

In another simulation, we kept the ratio of total removed mass to total added mass but removed the mass at a lower rate for a longer time: The duration of the accretion pulses and the mass addition rate are still $\Delta t_{\rm p}=0.1\yr$ and $\dot M_{\rm add}=0.03 M_\odot \yr^{-1}$, respectively, but the mass removal rate is $\dot M_{\rm rem}=-0.004 M_\odot \yr^{-1}$ during a time of  $5 \Delta t_{\rm p}=0.5\yr$ (for 168 pulses); namely, effectively we have $\eta_{\rm MR}=0.67$. The total cycle time is $0.1 + 0.5=0.6 \yr$, with a net addition of $0.001 M_\odot$. The net average mass accretion rate is $\dot M_{\rm acc} = 0.00167 M_\odot \yr^{-1}$.  

{ We note the results' sensitivity to the duration of the accretion and ejection phases.  The connection to the behavior of the accretion and jet-driven mass removal in real systems is through the operation of jets in a negative feedback cycle. The feedback cycle `chooses' the correct parameters. We discuss it here briefly, returning to the feedback cycle in Section \ref{sec:Summary}.  
As the star rapidly expands, the outer envelope layers engulf the inner parts of the accretion disk that launch jets inside the envelope. These jets remove the envelope's outskirts. The system reaches an equilibrium where the jets' power and accretion power balance each other to maintain a more or less constant stellar radius or oscillations around a slowly changing radius. The duration of the pulses we have in our numerical scheme is a numerical artifact. }

\begin{figure}[t]
	\centering
\includegraphics[trim=3.4cm 8.4cm 3.5cm 7.9cm ,clip, scale=0.62]{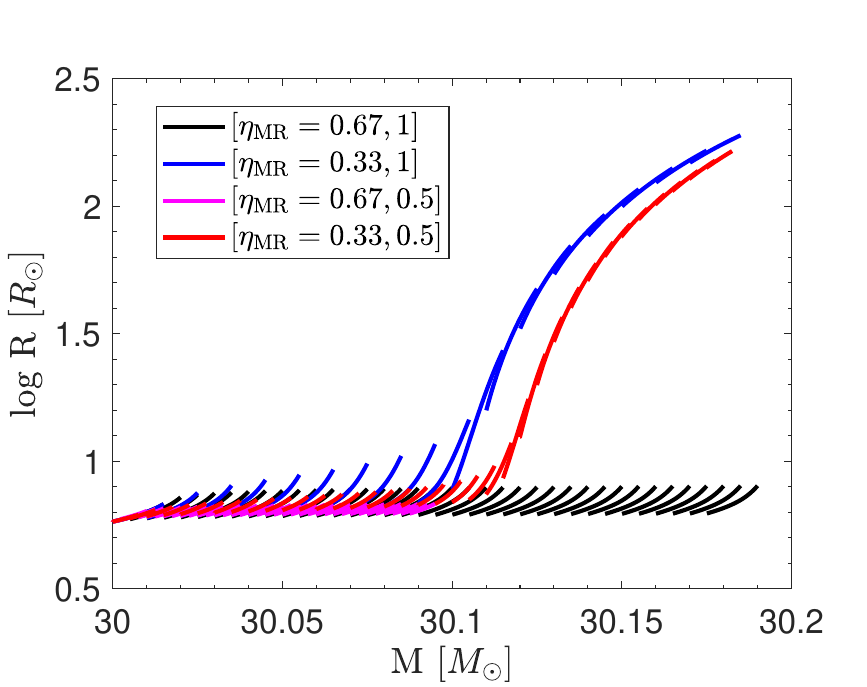}
\caption{Stellar radius (on a log scale) versus stellar mass, similar to \ref{fig:60Mo_logR_M} but for an initial mass of $M_{\rm ZAMS}=30M_\odot$. This image represents different accretion pulses. Black line (36 pulses) represents $\Delta t_p=1\yr$ and    mass removal at $\eta_{\rm MR} = \frac{2}{3}$.
Blue line (18 pulses) represents $\Delta t_{\rm p}=1\yr$ and mass removal at $\eta_{\rm MR} = \frac{1}{3}$. 
Magenta line (36 pulses) represents $\Delta t_{\rm p}=0.5\yr$ and  $\eta_{\rm MR} = \frac{2}{3}$.
Red line (36 pulses) represents $\Delta t_{\rm p}=0.5\yr$ for mass removal rate of $\eta_{\rm MR} = \frac{1}{3}$.
The accretion rate for all lines is $0.015 M_\odot \yr^{-1}$, which is calibrated by half of $0.03 M_\odot \yr^{-1}$ which was the accretion rate of $M_{\rm ZAMS}=60M_\odot$. 
 Energy is deposited along with the mass at a power of $\dot E_{\rm add} = 2.34\times 10^{39}$, which is $\eta_{\rm acc}=0.25$ of the accretion energy; the energy is deposited in the outer $10\%$ of the star (by radius).}
\label{fig:30Mo_logR_M}
\end{figure}

We also examine the response of a main sequence stellar model of $M_{\rm ZAMS}=30M_\odot$. We present the results of some cases in Figure \ref{fig:30Mo_logR_M}.
The mass addition rate during the mass addition phases (the pulses) is $\dot M_{\rm add} = 0.015 M_\odot \yr^{-1}$, half of that for the $M_{\rm ZAMS}=60M_\odot$ simulations.  
Figure \ref{fig:30Mo_logR_M} shows that when we remove enough mass in the mass removal phase of each cycle, namely, $\eta_{\rm MR} \ga 2/3$ (we did not scan the parameter space as we explain in Section  \ref{sec:Summary}), the star does not expand much. The black and magenta lines represent such cases. 
The red line represents the case with $\Delta t_{\rm p}=0.5\yr$ and a low value of $\eta_{\rm MR}=1/3$; expansion occurs later after more mass is accreted. 
For $\Delta t_{\rm p}=1 \yr$ and  $\eta_{\rm MR}=1/3$ that the blue line represents, expansion occurs earlier since insufficient mass is being removed.  

\begin{figure*}[t]
	\centering
 \includegraphics[trim=0.0cm 4.9cm 0.5cm 3.0cm ,clip, scale=0.8]{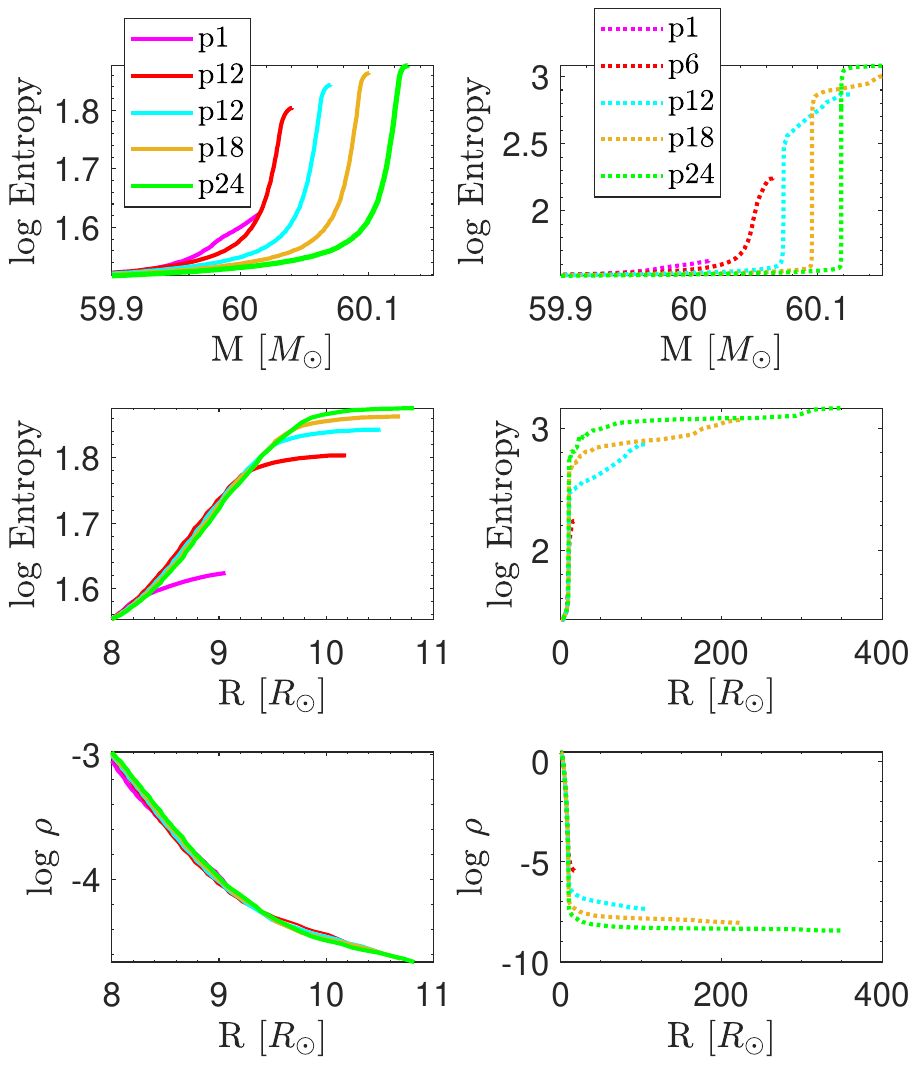}
\caption{Entropy (upper two rows) and density (lower row) profiles along several evolutionary points as we add mass, for two simulations with $M_{\rm ZAMS}=60M_\odot$ and $\Delta t_{\rm p}=0.5\yr$. Entropy is as given by \textsc{mesa}, which is the entropy per gram in units of $N_{\rm A} k_{\rm B}$ (the Avogadro number times Boltzmann constant).  
The three left panels are for the simulation with large removal mass $\eta_{\rm MR}=2/3$, and the three right panels (with dotted lines) are for a simulation with low removal mass $\eta_{\rm MR}=1/3$. The color of the lines is according to the pulse number given in each column's inset. }
\label{fig:60Mo_Entropy_r_profile}
\end{figure*}

\begin{figure*}[t]
	\centering
 \includegraphics[trim=0.0cm 4.9cm 0.5cm 3.0cm ,clip, scale=0.8]{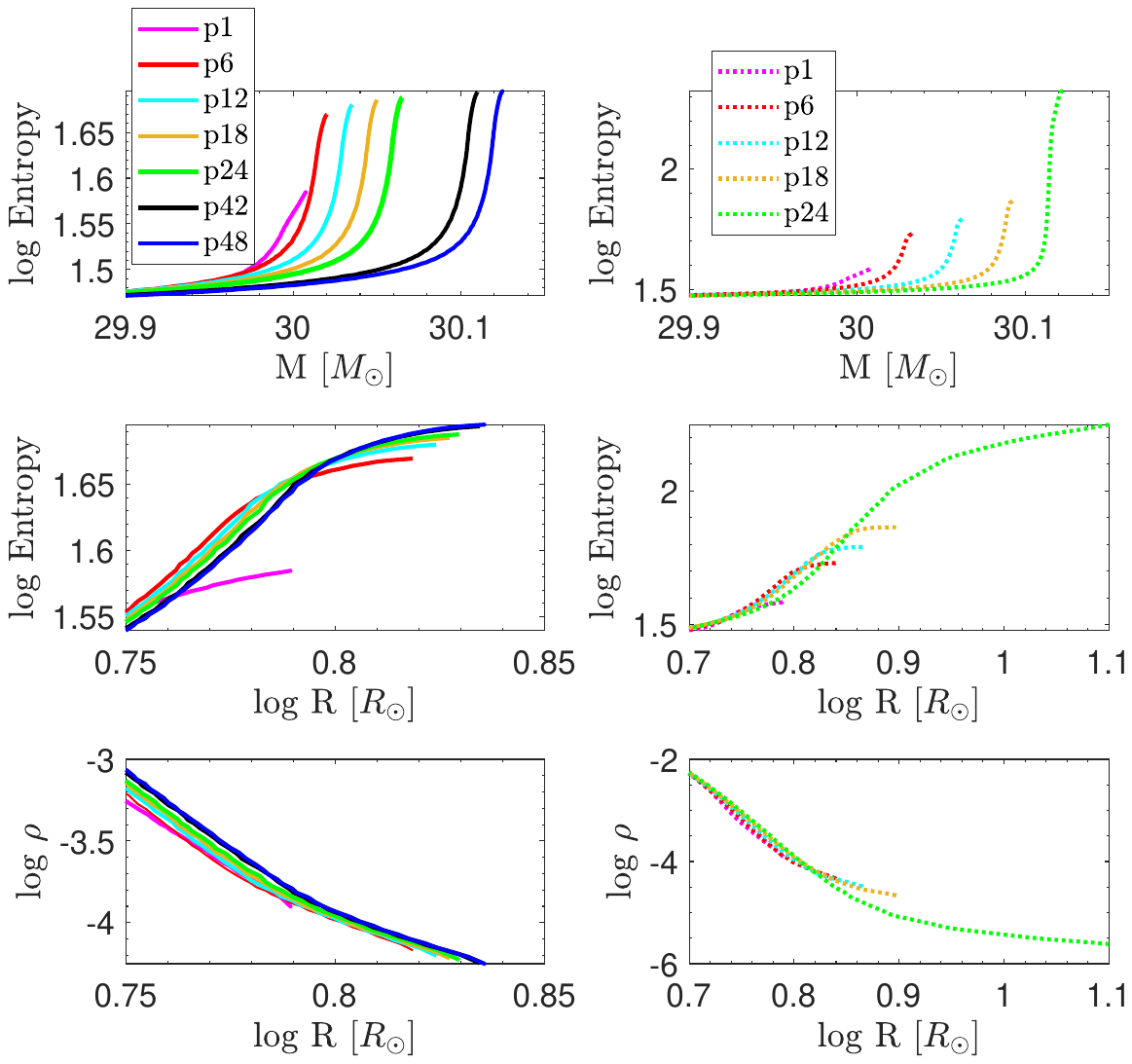}
\caption{Similar to Figure \ref{fig:60Mo_Entropy_r_profile}, but for $M_{\rm ZAMS}=30M_\odot$ and $\Delta t_{\rm p}=0.5\yr$. 
The left column is for $\eta_{\rm MR}=2/3$, 
and the right column is for $\eta_{\rm MR}=1/3$. 
}
\label{fig:30Mo_Entropy_r_profile}
\end{figure*}

In Figure \ref{fig:60Mo_Entropy_r_profile}, we present entropy and density profiles at several points as we add more mass to the $M_{\rm ZAMS}=60M_\odot$ stellar model. The insets give the pulse number of each profile. The first row shows the entropy profile given by MESA as a function of the mass coordinate in the outer region where we add mass, while the second row gives the entropy as a function of radius. The bottom row gives the density as a function of radius.  Figure \ref{fig:30Mo_Entropy_r_profile} presents similar profiles for two simulations of $M_{\rm ZAMS}=30M_\odot$.

The density profiles in the bottom row of Figures \ref{fig:60Mo_Entropy_r_profile} and \ref{fig:30Mo_Entropy_r_profile} show, as expected, the expansion of the envelope. In the case of large mass removal (left column), the star does not expand much. 

The entropy profiles, shown in the upper row as a function of mass, are important to our discussion. They show that as we add mass, a sharp entropy rise develops at the edge of the envelope, i.e., a sharp entropy rise in a very thin mass layer. If we remove this high-entropy thin mass layer, the star contracts. This explains why mass removal substantially reduces, or even prevents, the star's expansion. 
A star with a radiative envelope can grow in mass without expending much if the high-entropy outer layers are removed alongside the mass accretion. These outer layers have less mass than the added mass, so the net average mass accretion is positive.

\section{Discussion and Summary}
\label{sec:Summary}
 
We mimic a process where a very massive main sequence star accretes mass via an accretion disk, and this accretion disk launches energetic jets from its inner zones attached to the star. Namely, the jets carry most of the gravitational energy that the accreted mass releases. Moreover, we assume that the jets remove the outer layers of the mass-accreting star. This occurs because the star expands as it accretes mass at a high rate to radii larger than the inner radius of the accretion disk. The jets that the accretion disk launches from its inner zone collide with the rarefied outer layers of the swollen star and remove mass from them.  
We alternately added and removed mass from the stellar models to mimic this process. The numerical code \textsc{mesa} (Section \ref{sec:Method}) adds and removes mass from the outer layer. The accreted and removed mass has the same properties, like entropy, as the outermost layer. After a mass accretion episode (the pulse), the star expands, and the outermost layer has higher entropy than the former outermost layer (Figures \ref{fig:60Mo_Entropy_r_profile} and \ref{fig:30Mo_Entropy_r_profile}). In the next phase, we remove mass from the outer stellar layer; this layer has high entropy, so the star shrinks.  

We simulated the accretion process onto non-rotating spherically symmetric main sequence stellar models of $30 M_\odot$ and $60M_\odot$. We found (Section \ref{sec:Results}) that for not too short alternating inflow-outflow cycles and for an outflow that carries more than about half the accreted mass, the mass loss substantially reduces the rate of the expansion of the mass-accreting star (Figures \ref{fig:60Mo_logR_M} - \ref{fig:30Mo_logR_M}). We did not explore a large parameter space for two reasons. Firstly, the simulations take time, and we aim to show that substantial mass removal, as might occur when the accretion disk launches energetic jets, allows for a high mass accretion rate without much expansion. Secondly, mass removal by jets operates in a negative feedback cycle. If the star expands, the jets immediately remove more mass, so the star contracts. If the star contracts, the jets are less efficient in removing envelope mass, and the accretion leads to expansion. The feedback mechanism operates in a way that prevents much expansion (unless the mass accretion rate is much too high). 

\cite{SchurmannLanger2024} use \textsc{mesa} to thoroughly study main sequence stars accreting at constant rates. The parameter space they cover with their simulations is much larger than ours. They simulate only \textit{pure accretion}, i.e., accretion without mass loss. The main difference from our study is that they did not include mass loss during the mass-accretion process. This is a crucial difference in the accretion process and the results. They also did not inject energy alongside the accretion process, while we did. We differ on these ingredients also from the simulation that  \cite{Lauetal2024}, who simulated stellar models with masses of $M_{\rm ZAMS} \le 20 M_\odot$, lower than what we simulated here. 

We first compare here the results for $M=30 M_\odot$. 
\cite{SchurmannLanger2024} find that the general trend is that when the mass accretion timescale is shorter than the thermal timescale, the star expands. 
For the $M=30 M_\odot$ stellar model, \cite{SchurmannLanger2024} find that at an accretion rate of $\dot M_{\rm acc} = 3 \times 10^{-5} M_\odot \yr^{-1}$ the star expands by $30\%$ as its mass grows to $34 M_\odot$. For accretion rates of $\dot M_{\rm acc} \ga 10^{-4} M_\odot \yr^{-1}$ their stellar model expands unstably. We find, as an example, that when we add mass at a rate of $\dot M_{\rm acc}=0.015M_\odot \yr^{-1}$ in the mass addition part of the cycle and remove $2/3$ of this mass in the mass removal part, namely, a net accretion rate of $\dot M_{\rm acc} = 0.0025 M_\odot \yr^{-1}$, the stellar model does not expand much (black lines in Figure \ref{fig:30Mo_logR_M}). This holds even as we inject extra energy into the envelope with the parameter of $\eta_{\rm acc}=0.25$, i.e., a fraction of $0.25$ of the accretion energy is deposited to the envelope (the rest is carried by jets). 
{We obtain a very rapid radius expansion with a pure accretion (no mass removal) even if we add no energy. When we add $1.6 M_\odot$, the radius increases by a factor of 10. }

{ \cite{SchurmannLanger2024} do not simulate a star of $M=60 M_\odot$; their models of $M=50 M_\odot$ and $M=70 M_\odot$ expand very slowly when mass accretion rate is $\dot M_{\rm acc} = 10^{-4} M_\odot \yr^{-1}$. For accretion rates of $\dot M_{\rm acc} \ga 2 \times 10^{-4} M_\odot \yr^{-1}$ these two stellar model expand unstable. We find that under the right conditions, i.e., that we remove 2/3 of the mass we add in each pulse, our stellar model of $M=60 M_\odot$ does not expand much even at a net mass accretion rate as high as $\dot M_{\rm acc} = 5 \times 10^{-3} M_\odot \yr^{-1}$ (Figure \ref{fig:60Mo_logR_M}). }
These comparisons further emphasize the result that mass removal, even when we inject energy that acts to expand the star, prevents a large expansion. 

Our results are most important in vigorous binary interactions, where mass transfer at high rates leads the mass-accreting star to launch jets, e.g., the grazing envelope evolution. 
Such likely has been the case in the 1837-1856 Great Eruption of Eta Carinae (e.g., \citealt{Soker2001, AkashiKashi2020}). The companion mass in the jet-powered model of the Great Eruption is in the range of $M_2(\eta~Car) \simeq 30-80 M_\odot$, and more likely in the upper part of this range. Near periastron passages during the Great Eruption, the system experienced a grazing envelope evolution. The secondary star accreted $M_{\rm acc} \simeq 4 M_\odot$ during these 20 years \citep{KashiSoker2010}. Our results show that such a secondary can accrete mass at a high rate without expanding much if jets indeed remove mass from the outer high-entropy layers of the envelope. In that case, the gravitational potential well of the accreting star stays deep, and the jets are powerful. Our results allow for the high mass-accretion rate required by the jet-powered binary model of the Great Eruption. 
\footnote{The triple-star models of the Great Eruption \citep{PortegiesZwartvandenHeuvel2016, Hirai2021} suffer from difficulties, and we consider them unlikely. 
($i$) The Lesser Eruption in 1890-1895 (\citealt{Humphreysetal1999}) required another merging star according to the triple-star scenarios. Namely, a system of initial four stars. ($ii$) The Homunculus and the present binary system share an equatorial plane (e.g., \citealt{Maduraetal2012}). This requires a coplanar triple-stellar system; merger scenarios require an unstable triple system that will likely not be coplanar. ($iii$) The scenario by \cite{Hirai2021} predicts the presence of a dense gas in the equatorial plane; this is not observed in the Homunculus. Applying a triple-star or a quadruple-star model to the two nineteenth-century eruptions of Eta Carinae is unnecessary and problematic. }  
If this also holds for lower-mass stars (under study), our results allow high-mass-accretion rates in other types of ILOTs; some recent studies argue that only jets can power luminous ILOTs (e.g., \citealt{Soker2024}).


\section*{Acknowledgments}

{ We thank an anonymous referee for comments that improved the presentation of our results. }
A grant from the Pazy Foundation supported this research.



\label{lastpage}


\begin{thebibliography}{}\addcontentsline{toc}{section}{References}



\bibitem[Addison et al.(2022)]{Addisonetal2022} Addison H., Blagorodnova N., Groot P.~J., Erasmus N., Jones D., Mogawana O., 2022, MNRAS, 517, 1884. 

\bibitem[Akashi \& Kashi(2020)]{AkashiKashi2020} Akashi M., Kashi A., 2020, MNRAS, 494, 3186. 


\bibitem[Angulo et al.(1999)]{Anguloetal1999} Angulo, C., Arnould, M., Rayet, M., et al.\ 1999, \nphysa, 656, 3. 

\bibitem[Banerjee et al.(2020)]{Banerjeeetal2020} Banerjee, D.~P.~K., Geballe, T.~R., Evans, A., Shahbandeh, M., Woodward, C.~E., Gehrz, R.~D., Eyres, S.~P.~S., Starrfield, S. \& Zijlstra, A.\ 2020, \apjl, 904, L23. 

\bibitem[Berger et al.(2009)]{Berger2009} Berger, E., Soderberg, A. M., Chevalier, R. A., et al. 2009, \apj, 699, 1850

\bibitem[Blagorodnova et al.(2020)]{Blagorodnovaetal2020} Blagorodnova, N., Karambelkar, V., Adams, S.~M., et al.\ 2020, \mnras, 496, 5503. doi:10.1093/mnras/staa1872 

\bibitem[\protect\citeauthoryear{Blackman \& Lucchini}{2014}]{BlackmanLucchini2014} Blackman E.~G., Lucchini S., 2014, MNRAS, 440, L16. doi:10.1093/mnrasl/slu001

\bibitem[Blagorodnova et al.(2021)]{Blagorodnovaetal2021} Blagorodnova N., Klencki J., Pejcha O., Vreeswijk P.~M., Bond H.~E., Burdge K.~B., De K., et al., 2021, A\&A, 653, A134. doi:10.1051/0004-6361/202140525

\bibitem[Blagorodnova et al.(2017)]{Blagorodnovaetal2017} Blagorodnova, N., Kotak, R., Polshaw, J., et al.\ 2017, \apj, 834, 107

\bibitem[Blouin et al.(2020)]{Blouinetal2020} Blouin, S., Shaffer, N.~R., Saumon, D., \&  Starrett, C. E. \ 2020, \apj, 899, 46. 

\bibitem[Boian \& Groh(2019)]{BoianGroh2019} Boian, I., \& Groh, J.~H.\ 2019, \aap, 621, A109.

\bibitem[Bond et al.(2003)]{Bondetal2003} Bond, H.~E., Henden, A., Levay, Z.~G., et al.\ 2003, \nat, 422, 405

\bibitem[Cai et al.(2019)]{Caietal2019} Cai Y.-Z., Pastorello A., Fraser M., Prentice S.~J., Reynolds T.~M., Cappellaro E., Benetti S., et al., 2019, A\&A, 632, L6. 

\bibitem[Cai et al.(2022a)]{Caietal2022} Cai Y.-Z., Pastorello A., Fraser M., Wang X.-F., Filippenko A.~V., Reguitti A., Patra K.~C., et al., 2022a, A\&A, 667, A4. 

\bibitem[Cai et al.(2022b)]{Caietal2022b} Cai Y., Reguitti A., Valerin G., Wang X., 2022b, Univ, 8, 493. 

\bibitem[Cassisi et al.(2007)]{Cassisietal2007} Cassisi, S., Potekhin, A.~Y., Pietrinferni, A.,  Catelan, M., \&  Salaris, M. \ 2007, \apj, 661, 1094. 

\bibitem[Chugunov et al.(2007)]{Chugunovetal2007} Chugunov, A.~I., Dewitt, H.~E., \& Yakovlev, D.~G.\ 2007, \prd, 76, 025028. 

\bibitem[Cyburt et al.(2010)]{Cyburtetal2010} Cyburt, R.~H., Amthor, A.~M., Ferguson, R., et al.\ 2010, \apjs, 189, 240. 

\bibitem[Danieli \& Soker(2019)]{DanieliSoker2019} Danieli B., Soker N., 2019, MNRAS, 482, 2277. 


\bibitem[De et al.(2023)]{Deetal2023} De K., MacLeod M., Karambelkar V., Jencson J.~E., Chakrabarty D., Conroy C., Dekany R., et al., 2023, Natur, 617, 55.  

\bibitem[Ferguson et al.(2005)]{Fergusonetal2005} Ferguson, J.~W., Alexander, D.~R., Allard, F., Barman, T., Bodnarik, J. G., Hauschildt, P. H., Heffner-Wong, A., \& Tamanai, A.\ 2005, \apj, 623, 585. 

\bibitem[Fuller et al.(1985)]{Fulleretal1985} Fuller, G.~M., Fowler, W.~A., \& Newman, M.~J.\ 1985, \apj, 293, 1. 

\bibitem[Gilkis et al.(2019)]{Gilkisetal2019} Gilkis, A., Soker, N., \& Kashi, A.\ 2019, \mnras, 482, 4233

\bibitem[Grichener \& Soker(2019)]{GrichenerSoker2019} Grichener A., Soker N., 2019, ApJ, 878, 24. 


\bibitem[Gurevich, Bear, \& Soker(2022)]{Gurevichetal2022} Gurevich O., Bear E., Soker N., 2022, MNRAS, 511, 1330. 

\bibitem[Hirai et al.(2021)]{Hirai2021} Hirai R., Podsiadlowski P., Owocki S.~P., Schneider F.~R.~N., Smith N., 2021, MNRAS, 503, 4276. 

\bibitem[Howitt et al.(2020)]{Howittetal2020} Howitt, G., Stevenson, S., Vigna-G{\'o}mez, A., et al.\ 2020, \mnras, 492, 3229

\bibitem[Hubov{\'a}, \& Pejcha(2019)]{HubovaPejcha2019} Hubov{\'a}, D., \& Pejcha, O.\ 2019, \mnras, 489, 891

\bibitem[Humphreys, Davidson, \& Smith(1999)]{Humphreysetal1999} Humphreys R.~M., Davidson K., Smith N., 1999, PASP, 111, 1124. 

\bibitem[Iglesias \& Rogers(1993)]{IglesiasRogers1993} Iglesias, C.~A. \& Rogers, F.~J.\ 1993, \apj, 412, 752. 

\bibitem[Iglesias \& Rogers(1996)]{IglesiasRogers1996} Iglesias, C.~A. \& Rogers, F.~J.\ 1996, \apj, 464, 943. 

\bibitem[Irwin (2004)] {Irwin2004} Irwin, Alan. W. (2004). The FreeEOS Code for Calculating the Equation of State for Stellar
Interiors (cit. on p. xxi).

\bibitem[Itoh et al.(1996)]{Itohetal1996} Itoh, N., Hayashi, H., Nishikawa, A., Kohyama, Y. \ 1996, \apjs, 102, 411. 

\bibitem[Ivanova et al.(2013)]{Ivanovaetal2013a} Ivanova, N., Justham, S., Avendano Nandez, J.~L., \& Lombardi, J.~C.\ 2013, Science, 339, 433

\bibitem[Jencson et al.(2019)]{Jencsonetal2019} Jencson, J.~E., Kasliwal, M.~M., Adams, S.~M., et al.\ 2019, \apj, 886, 40

\bibitem[Jermyn et al.(2021)]{Jermynetal2021} Jermyn, A.~S., Schwab, J., Bauer, E., Timmes F. X., \& Potekhin A. Y.\ 2021, \apj, 913, 72. 


\bibitem[Jones(2020)]{Jones2020} Jones, D.\ 2020, Reviews in Frontiers of Modern Astrophysics; From Space Debris to Cosmology, 123. 

\bibitem[Kaminski(2024)]{Kaminski2024} Kaminski T., 2024, arXiv, arXiv:2401.03919. 


\bibitem[Kami{\'n}ski et al.(2015)]{Kaminskietal2015} Kami{\'n}ski, T., Mason, E., Tylenda, R., \& Schmidt, M.~R.\ 2015, \aap, 580, A34

\bibitem[Kami{\'n}ski et al.(2020)]{Kaminskietal2020Nova1670} Kami{\'n}ski, T., Menten, K.~M., Tylenda, R., et al.\ 2020, \aap, 644, A59. doi:10.1051/0004-6361/202038648 

\bibitem[Kami{\'n}ski et al.(2023)]{Kaminskietal2023} Kami{\'n}ski T., Schmidt M., Hajduk M., Kiljan A., Izviekova I., Frankowski A., 2023, A\&A, 672, A196. 

\bibitem[Kami{\'n}ski et al.(2021)]{Kaminskietal2021Nova1670} Kami{\'n}ski, T., Steffen, W., Bujarrabal, V., et al.\ 2021, \aap, 646, A1. doi:10.1051/0004-6361/202039634 

\bibitem[Kaminski et al.(2018)]{Kaminskietal2018} Kaminski, T., Steffen, W., Tylenda, R.,  Young, K.~H., Patel, N.~A., \& Menten, K.~M.\ 2018, \aap, 617, A129


\bibitem[Karambelkar et al.(2023)]{Karambelkaretal2023} Karambelkar V.~R., Kasliwal M.~M., Blagorodnova N., Sollerman J., Aloisi R., Anand S.~G., Andreoni I., et al., 2023, ApJ, 948, 137. 


\bibitem[Kashi et al.(2019)]{Kashietal2019Galax} Kashi, A., Michaelis, A.~M., \& Feigin, L.\ 2019, Galaxies, 8, 2. 

\bibitem[Kashi \& Soker(2010)]{KashiSoker2010} Kashi A., Soker N., 2010, ApJ, 723, 602. 

\bibitem[Kashi \& Soker(2016a)]{KashiSoker2016Terms} Kashi A., Soker N., 2016a, RAA, 16, 99. doi:10.1088/1674-4527/16/6/099

\bibitem[Kashi \& Soker(2016b)]{KashiSoker2016EtaCar} Kashi A., Soker N., 2016b, ApJ, 825, 105. 


\bibitem[Kasliwal(2011)]{Kasliwal2011} Kasliwal, M.~M.\ 2011, Bulletin of the Astronomical Society of India, 39, 375

\bibitem[Kasliwal et al.(2012)]{Kasliwaletal2012} Kasliwal, M.~M., Kulkarni, S.~R., Gal-Yam, A., et al.\ 2012, \apj, 755, 161


\bibitem[Klencki et al.(2021)]{Klenckietal2021} Klencki, J., Nelemans, G., Istrate, A.~G., \& Chruslinska, M., \ 2021, \aap, 645, A54. 

\bibitem[Jermyn et al.(2023)]{Jermynetal2023} Jermyn, A.~S., Bauer, E.~B., Schwab, J., et al.\ 2023, \apjs, 265, 15. 

\bibitem[Langanke \& Mart{\'\i}nez-Pinedo(2000)]{Langankeetal2000} Langanke, K. \& Mart{\'\i}nez-Pinedo, G.\ 2000, \nphysa, 673, 481. 

\bibitem[\protect\citeauthoryear{Lau et al.}{2024}]{Lauetal2024} Lau M.~Y.~M., Hirai R., Mandel I., Tout C.~A., 2024, ApJL, 966, L7. 


\bibitem[MacLeod \& Loeb(2020)]{MacLeodLoeb2020} MacLeod, M. \& Loeb, A.\ 2020, \apj, 895, 29. 

\bibitem[MacLeod et al.(2017)]{MacLeodetal2017} MacLeod, M., Macias, P., Ramirez-Ruiz, E., Grindlay, J., Batta, A., \& Montes, G.\ 2017, \apj, 835, 282
   
\bibitem[MacLeod et al.(2018)]{MacLeodetal2018} MacLeod, M., Ostriker, E.~C., \& Stone, J.~M.\ 2018, \apj, 868, 136.

\bibitem[Madura et al.(2012)]{Maduraetal2012} Madura T.~I., Gull T.~R., Owocki S.~P., Groh J.~H., Okazaki A.~T., Russell C.~M.~P., 2012, MNRAS, 420, 2064. 

\bibitem[Mason et al.(2010)]{Masonetal2010} Mason, E., Diaz, M., Williams, R.~E., Preston, G., \& Bensby, T.\ 2010, \aap, 516, A108

\bibitem[Mcley \& Soker(2014)]{McleySoker2014} Mcley L., Soker N., 2014, MNRAS, 445, 2492. 

\bibitem[Matsumoto \& Metzger(2022)]{MatsumotoMetzger2022} Matsumoto T., Metzger B.~D., 2022, ApJ, 938, 5. 

\bibitem[Metzger, Giannios, \& Spiegel(2012)]{Metzgeretal2012} Metzger B.~D., Giannios D., Spiegel D.~S., 2012, MNRAS, 425, 2778. 

\bibitem[Metzger \& Pejcha(2017)]{MetzgerPejcha2017} Metzger, B.~D., \& Pejcha, O.\ 2017, \mnras, 471, 3200
 

\bibitem[Mobeen et al.(2021)]{Mobeenetal2021} Mobeen M.~Z., Kami{\'n}ski T., Matter A., Wittkowski M., Paladini C., 2021, A\&A, 655, A100. 

\bibitem[Mould et al.(1990)]{Mouldetal1990} Mould, J., Cohen, J., Graham, J.~R., et al.\ 1990, \apjl, 353, L35


\bibitem[Muthukrishna et al.(2019)]{MuthukrishnaetalM2019} Muthukrishna, D., Narayan, G., Mandel, K.~S., Biswas, R., \& Hlo{\v z}ek, R.\ 2019, \pasp, 131, 118002
  
\bibitem[Nandez et al.(2014)]{Nandezetal2014} Nandez, J.~L.~A., Ivanova, N., \& Lombardi, J.~C., Jr.\ 2014, \apj, 786, 39

\bibitem[O'Connor et al.(2023)]{Oconnoretal2023} O'Connor C.~E., Bildsten L., Cantiello M., Lai D., 2023, ApJ, 950, 128. doi:10.3847/1538-4357/acd2d4

\bibitem[Oda et al.(1994)]{Odaetal1994} Oda, T., Hino, M., Muto, K., Takahara M., \& Sato K.\ 1994, Atomic Data and Nuclear Data Tables, 56, 231. 

\bibitem[Ofek et al.(2008)]{Ofek2008} Ofek, E.~O., Kulkarni, S.~R., Rau, A., et al.\ 2008, \apj, 674, 447

\bibitem[Pastorello \& Fraser(2019)]{PastorelloFraser2019} Pastorello, A., \& Fraser, M.\ 2019, Nature Astronomy, 3, 676


\bibitem[Pastorello et al.(2019)]{PastorelloMasonetal2019} Pastorello, A., Mason, E., Taubenberger, S., et al.\ 2019, \aap, 630, A75

\bibitem[Pastorello et al.(2021)]{Pastorelloetal2021} Pastorello A., Valerin G., Fraser M., Elias-Rosa N., Valenti S., Reguitti A., Mazzali P.~A., et al., 2021, A\&A, 647, A93. 

\bibitem[Pastorello et al.(2023)]{Pastorelloetal2023} Pastorello A., Valerin G., Fraser M., Reguitti A., Elias-Rosa N., Filippenko A.~V., Rojas-Bravo C., et al., 2023, A\&A, 671, A158. 

\bibitem[Paxton et al.(2011)]{Paxtonetal2011} Paxton, B., Bildsten, L., Dotter, A., et al.\ 2011, \apjs, 192, 3

\bibitem[Paxton et al.(2013)] {Paxtonetal2013} Paxton, B., Cantiello, M., Arras, P., et al. 2013, \apjs, 208, 4

\bibitem[Paxton et al.(2015)]{Paxtonetal2015} Paxton, B., Marchant, P., Schwab, J., et al.\ 2015, \apjs, 220, 15

\bibitem[Paxton et al.(2018)]{Paxtonetal2018} Paxton, B., Schwab, J., Bauer, E.~B., et al.\ 2018, \apjs, 234, 34

\bibitem[Paxton et al.(2019)]{Paxtonetal2019} Paxton, B., Smolec, R., Schwab, J., et al.\ 2019, \apjs, 243, 10, 

\bibitem[Pejcha et al.(2016a)]{Pejchaetal2016a} Pejcha, O., Metzger, B.~D., \& Tomida, K.\ 2016a, \mnras, 455, 4351
  
\bibitem[Pejcha et al.(2016b)]{Pejchaetal2016b} Pejcha, O., Metzger, B.~D., \& Tomida, K.\ 2016b, \mnras, 461, 2527

\bibitem[Pejcha et al.(2017)]{Pejchaetal2017} Pejcha O., Metzger B.~D., Tyles J.~G., Tomida K., 2017, ApJ, 850, 59. 

\bibitem[Portegies Zwart \& van den Heuvel(2016)]{PortegiesZwartvandenHeuvel2016} Portegies Zwart S.~F., van den Heuvel E.~P.~J., 2016, MNRAS, 456, 3401. 

\bibitem[Potekhin \& Chabrier(2010)]{PotekhinChabrier2010} Potekhin, A.~Y. \& Chabrier, G.\ 2010, Contributions to Plasma Physics, 50, 82. 

\bibitem[Poutanen(2017)]{Poutanen2017} Poutanen, J.\ 2017, \apj, 835, 119. 

\bibitem[Qian et al.(2020)]{Qianetal2020} Qian S.-B., Zhu L.-Y., Liu L., Zhang X.-D., Shi X.-D., He J.-J., Zhang J., 2020, RAA, 20, 163. 

\bibitem[Rau et al.(2007)]{Rau2007} Rau, A., Kulkarni, S.~R., Ofek, E.~O., \& Yan, L.\ 2007, \apj, 659, 1536

\bibitem[Retter \& Marom(2003)]{RetterMarom2003} Retter, A., \& Marom, A.\ 2003, \mnras, 345, L25

\bibitem[Rogers \& Nayfonov(2002)]{RogersNayfonov2002} Rogers, F.~J. \& Nayfonov, A.\ 2002, \apj, 576, 1064. 

\bibitem[Saumon et al.(1995)]{Saumonetal1995} Saumon, D., Chabrier, G., \& van Horn, H.~M.\ 1995, \apjs, 99, 713. 


\bibitem[Schr{\o}der et al.(2020)]{Schrderetal2020} Schr{\o}der, S.~L., MacLeod, M., Loeb, A., et al.\ 2020, \apj, 892, 13. doi:10.3847/1538-4357/ab7014 

\bibitem[Schreier et al.(2021)]{Schreieretal2021} Schreier, R., Hillel, S., Shiber, S., \& Soker, N.\ 2021, \mnras. 

\bibitem[Sch{\"u}rmann \& Langer(2024)]{SchurmannLanger2024} Sch{\"u}rmann C., Langer N., 2024, A\&A, 691, A174. 


\bibitem[Segev et al.(2019)]{Segevetal2019} Segev, R., Sabach, E., \& Soker, N.\ 2019, \apj, 884, 58


\bibitem[Soker(2001)]{Soker2001} Soker N., 2001, MNRAS, 325, 584. 


\bibitem[Soker(2016a)]{Soker2016GEE} Soker N., 2016a, NewA, 47, 16. 

\bibitem[Soker(2020)]{Soker2020ILOTjets} Soker N., 2020, ApJ, 893, 20. 
 

\bibitem[Soker(2022)]{Soker2022Rev} Soker, N., 2022,  Research in Astronomy and Astrophysics, 22, 122003. 

\bibitem[Soker(2023)]{Soker2023BrightILOT} Soker N., 2023, OJAp, 6, 32. 

\bibitem[Soker(2024)]{Soker2024} Soker N., 2024, arXiv:2404.19617. 

\bibitem[Soker \& Gilkis(2018)]{SokerGilkis2018} Soker, N., \& Gilkis, A.\ 2018, \mnras, 475, 1198

\bibitem[Soker \& Kaplan(2021)]{SokerKaplan2021RAA} Soker N., Kaplan N., 2021, RAA, 21, 090. 

\bibitem[Stritzinger et al.(2020a)]{Stritzingeretal2020AT2014ej} Stritzinger M.~D., Taddia F., Fraser M., Tauris T.~M., Contreras C., Drybye S., Galbany L., et al., 2020a, A\&A, 639, A104. 

\bibitem[Stritzinger et al.(2020b)]{Stritzingeretal2020SNhunt120} Stritzinger M.~D., Taddia F., Fraser M., Tauris T.~M., Suntzeff N.~B., Contreras C., Drybye S., et al., 2020b, A\&A, 639, A103. 

\bibitem[Timmes \& Swesty(2000)]{TimmesSwesty2000} Timmes, F.~X. \& Swesty, F.~D.\ 2000, \apjs, 126, 501. 



\bibitem[Tsuna et al.(2024)]{Tsunaetal2024} Tsuna, D., Matsumoto, T., Wu, S.~C., Fuller J., 2024, \apj, 966, 30. doi:10.3847/1538-4357/ad3637

\bibitem[Tylenda et al.(2011)]{Tylendaetal2011} Tylenda, R., Hajduk, M., Kami{\'n}ski, T., et al.\ 2011, \aap, 528, A114


\bibitem[Tylenda et al.(2024)]{Tylendaetal2024} Tylenda, R., Kami{\'n}ski, T., \& Smolec, R.\ 2024, \aap, 685, A49. 

\bibitem[Tylenda et al.(2013)]{Tylendaetal2013} Tylenda, R., Kami{\'n}ski, T., Udalski, A., et al.\ 2013, \aap, 555, A16
    

\bibitem[Wadhwa et al.(2022)]{Wadhwaetal2022} Wadhwa S.~S., De Horta A., Filipovi{\'c} M.~D., Tothill N.~F.~H., Arbutina B., Petrovi{\'c} J., Djura{\v{s}}evi{\'c} G., 2022, RAA, 22, 105009. 


\bibitem[Yalinewich \& Matzner(2019)]{YalinewichMatzner2019} Yalinewich, A., \& Matzner, C.~D.\ 2019, \mnras, 490, 312

\bibitem[Yamazaki, Hayasaki, \& Loeb(2017)]{Yamazakietal2017} Yamazaki R., Hayasaki K., Loeb A., 2017, MNRAS, 466, 1421. 


\bibitem[Mobeen et al.(2024)]{ZainMobeenetal2024} Mobeen, M.~Z., Kami{\'n}ski, T., Matter, A., Wittkowski M., Monnier J.~D., Kraus S., Le Bouquin J.-B., et al.,\ 2024, \aap, 686, A260. doi:10.1051/0004-6361/202347322 



\bibitem[Zhu et al.(2023)]{Zhuetal2023} Zhu C.-H., L{\"u} G.-L., Lu X.-Z., He J., 2023, RAA, 23, 025021. 

\end{thebibliography}
\end{document}